\documentclass[usenatbib]{mn2e}
\usepackage{graphicx}
\usepackage{graphics}

\def\msun{{\rm M_{\odot}}}

\def\xte{{\it RXTE}}

\def\chisq{{\chi^{2}}}

\def\H0{{\rm ~km~s^{-1}~Mpc^{-1}}}

\def\delchi{\hbox{$\Delta\chi$}}
\def\msun{M_{\rm \odot}}

\def\gammamax{$\gamma_{max}$}
\def\Gammainj{$\Gamma_{inj}$}

\def\deg{^\circ}

\def\ls{$l_{\rm s}$}
\def\lh{$l_{\rm h}$}
\def\Ls{$L_{\rm s}$}
\def\Lh{$L_{\rm h}$}
\def\lth{$l_{\rm th}$}
\def\lnth{$l_{\rm nth}$}
\def\Lth{$L_{\rm th}$}

\begin{document}

\title[Pair production and multi-zone comptonization]{Limits on pair production and multi-zone comptonization\\ 
\Large {The broadband X/$\gamma$-ray spectrum of XTE J1550-564 revisited}}

\author[Hjalmarsdotter, Axelsson \& Done]{Linnea Hjalmarsdotter$^{1}$\thanks{email: astrogirl@telia.com},
Magnus Axelsson$^{2,3}$ and Chris Done$^{4}$\\
$^{1}$Sternberg Astronomical Institute, Lomonosov Moscow State University, Universitetskij pr. 13, 119991 Moscow, Russia\\
$^{2}$Department of Astronomy, Stockholm University, SE-106 91 Stockholm, Sweden\\
$^{3}$Oskar Klein Center for CosmoParticle Physics, Department of Physics, AlbaNova, SE-106 91 Stockholm, Sweden\\
$^{4}$Department of Physics, Durham University, South Road, Durham DH1 3LE, UK\\
}

\date{Accepted --. Received --; in original form --}

\pagerange{\pageref{firstpage}--\pageref{lastpage}} \pubyear{2013}

\maketitle

\begin{abstract}
At high luminosities black hole binaries show spectra with a strong disc component
accompanied by an equally strong tail where at least some of the electrons are non-thermal. We reanalyze the simultaneous ASCA-RXTE-OSSE data from the 1998 outburst of XTE J1550-564, which span 0.7-1000 keV and
remain the best data available of a black hole binary in this state. We reassess the importance of electron-positron pair production using a realistically high value of the source compactness for the first time. The lack of an observable annihilation line together with the observed $\gamma$-ray flux beyond 511 keV constrains the maximum electron Lorentz factor to be $\leq$10 and the slope of the injected electrons to $\leq$2.5. We also use the fast (10-50 Hz) variability spectrum to constrain the spatial dependence of the electron heating and acceleration. We find that the spectrum of the fast variability is consistent with being fully thermal, so that the observed non-thermal emission is produced from further out in the flow.

\end{abstract}

\section{Introduction}
The so-called very high state is perhaps the most complex and least understood spectral state of black hole binaries. It is a highly luminous state characterized by a strong blackbody component together with a very strong rather steep tail, typically starting at the peak of the disc component which can make these two components hard to separate from each other. The question of whether the disc in the very high state is truncated or not is a matter of debate. It is also not clear whether the strong signs of non-thermal Compton emission is the result of scattering off one single hybrid electron distribution, or if the thermal and non-thermal electrons come from different spatially separated regions of the flow. A popular idea of the possible configuration is that the thermal electrons are located close to the black hole in a region which may be the remains of the hot inner flow seen in the low/hard state, while the non-thermal electrons are located further out in the flow, perhaps accelerated by magnetic flares above the disc. To distinguish between a single hybrid distribution and a thermal plus a non-thermal (or to be correct, one with purely non-thermal injection, see discussion in Section 3) is not trivial but is of important for understanding the heating/acceleration processes involved. 

Further, the persistent non detection of any annihilation lines from Galactic black holes questions the importance of pair production in these sources, and is often used for assuming a strong upper limit on their compactness. While this may be justifiable in less luminous states, it is difficult to motivate in the very high state.

\subsection{XTE~J1550-564}
XTE~J1550-564 is a transient low-mass X-ray binary. It was discovered during an outburst in 1998 by the {\xte} All-Sky Monitor \citep[ASM;][]{smi98} and the {\it CGRO} Burst and Transient Source Experiment \citep[BATSE;][]{wil98}. The companion is a main sequence star of type G-K and the mass of the compact object has been estimated to 8.4-11.2 $\msun$ (Orosz02), and it is thus considered to be a black hole. The source displays all the canonical spectral states typically associated with black hole binaries. The X-ray spectrum and its evolution has been studied extensively since its discovery.

In this paper we reanalyse the simultaneous ASCA-RXTE-OSSE data from the 1998 outburst which spans 0.7-1000~keV and remain the best broadband data of an X-ray binary in the very high state. Previous modeling of the spectrum has not been able to utilise it to the fullest extent due to issues with the ASCA-RXTE cross-calibration and lack of good ionized reflection models. For the first time we now fit the entire broadband 0.7--1000 keV spectrum simultaneously. 

We use additional information from the fast (10--50 Hz) variability in the form of frequency resolved spectroscopy. This is a very powerful tool to help breaking the inherent degeneracy of spectral modelling, used e.g. in the pioneering work by \citet{churazov}. The technique has however not been used much for this purpose. Recently, \citet{AHD13}, hereafter AHD13, studied the PCA spectrum of XTE J1550-564 during the rise of the strong flare in 1998 and compared the spectrum of the continuum to that of the fast (10--50 Hz) variability. The difference that was found between the time averaged spectrum and the spectrum of the fast variability strongly supports spatially inhomogeneous comptonization models where (at least) two electron distributions seem to be present at different distances from the black hole. In this paper we develop this idea further and also attempt to separate the thermal and non-thermal distributions. 

The broadband data give us a good estimate of the source luminosity and compactness, which is high. This should lead to a non-negligible production of electron-positron pairs. The lack of an annihilation line in the OSSE data allows us to derive limits on pair production in this source that may be applicable to luminous black hole binaries in general.

\section{Data}
\subsection{PCA data continuum and frequency resolved spectra}
We use the data from the plateau phase, approximately two weeks after the initial flare.
For the PCA continuum data we extracted Standard2 spectra for all 8 observations between September 23 and October 6, applying standard selection criteria. As shown in \citet{gd03}, hereafter GD03 (their fig. 2), the position of the source in a PCA colour-colour diagram during this time corresponds to the very high spectral state. (This state was also labelled the extreme very high state by \citealt{kd04}, hereafter KD04, to separate it from a more disc dominated very high state also displayed by this source.) These were then co-added to create a total PCA spectrum with a total exposure time of 31 ks. A systematic error of 1 per cent was added to each bin. The energy band used for spectral modelling is 3--20 keV. 
To extract the spectrum of the fast variability we follow the procedure described in \citet{rev99} and \citet{rev01}. We extract a light curve for each available channel for each observation and construct a power density spectrum (PDS). The relative contribution of each channel was determined by integrating the PDS above 10 Hz, and used to construct an energy spectrum of the rapid variability. As with the continuum spectra, results from all observations were then co-added into a single spectrum, and a systematic error of one per cent added to each bin. We choose the frequency range 10--50 Hz in order to avoid contamination from the observed QPO at $\leq 4$ Hz and its harmonic (see further discussion in AHD13 and \citealt{adh14}.

\subsection{HEXTE and OSSE data}
We use the HEXTE data, averaged over the same period, from cluster 0 only. DG03 found the HEXTE data to be consistent with remaining constant throughout the period. 

The OSSE data are from viewing period 729.5, from September 25 to October 6, 1998. We use the same high-level product spectrum as in DG03, extracted for detectors 1,2,3 and 4 for the whole viewing period. 
The OSSE and HEXTE data are in very good agreement in the overlapping region between 50 and $\sim$120 keV above which the HEXTE data start to slightly flatten out. The agreement is however still within the HEXTE errors up to 200 keV. The OSSE data show no evidence of a high-energy cutoff up to at least 1000 keV. There is no sign of any annihilation line at 511 keV in the data.

Unfortunately the low sensitivity of HEXTE and low time resolution with retained spectral resolution of the OSSE data do not allow us to extend the fourier-spectrum to higher energies.  

\subsection{ASCA data}
The ASCA data used here are from a 24 ksec pointing on September 24, 1998 and is the same data as used in GD03, extending from 0.7--10 keV. In GD03 the ASCA data was treated separately to avoid previous cross-calibration issues with {\it RXTE}. These cross-calibration issues were fixed in \textsc{HEASOFT} 5.2 (and in KD04 this ASCA data were modelled together with some of the PCA and HEXTE pointings included in our co-added spectrum). The data were also used e.g. by \citet{steiner} (see discussion in Section 5).

\section{Spectral model}
\subsection{\textsc{Eqpair}}
We use the comptonization code \textsc{eqpair} \citet{eqpair}. The key-concept in this model is the compactness, defined as: 
 
\begin{equation}
l = \frac{L}{R} \frac{\sigma_{\rm T}}{m_{\rm e} c^{3}}
\end{equation}
where $L$ is the luminosity or power, $R$ the size of the comptonizing region, $\sigma_{\rm T }$ the Thomson cross section and $m_{\rm e}$ the electron mass.

The hard compactness, \lh, is determined by \Lh, the luminosity or power supplied to the electrons in the comptonizing region. This power can be supplied either as thermal heating \Lth, or as energetic electrons (and/or positrons) being injected into the source with a fixed non-thermal energy spectrum $Q(\gamma)$ that can be either mono-energetic or a power law with index {\Gammainj} extending from Lorentz factor $\gamma_{min}\sim$ 1--2 to $\gamma_{max}\sim$ 3--1000. The total hard compactness is thus \lh=\lth+\lnth. The ratio \lnth/\lh\ gives the relative importance of non-thermal scattering and if this parameter is $>0$ we have a hybrid electron distribution with a low-energy Maxwellian and a high-energy power law part. 

The soft compactness, \ls, is in turn determined by \Ls, the luminosity of soft photons available for Compton up-scattering and thus cooling of the hot electrons. If electron heating/acceleration dominates over cooling we get a hard spectrum and if cooling by soft photons dominates the resulting spectrum is soft, and the main parameter of the model determining the overall spectral state is thus the ratio \lh/\ls. The absolute value of the soft compactness \ls\ is also a fit parameter. This parameter not only measures the effectiveness of Compton cooling, compared to Coulomb cooling, but also the optical depth for pair-production in the source. Note that in \textsc{eqpair} \ls\ is not coupled to the normalization. In practice, the fit is therefore insensitive to this parameter (except regarding the strength of the annihilation line, see further discussion in the next section) and it is therefore often kept frozen to some fiducial value in the fitting process. The exact spectral shape for a given amount of seed photons is determined by a combination of the electron temperature and optical depth of the hot flow. The total optical depth $\tau_{T}$ may in addition to the electron optical depth $\tau_{p}$, which is a fit parameter, also include a part from pairs. Together with the electron temperature $kT_{e}$, $\tau_{T}$ is calculated by the model self consistently. The model does not automatically assume pair balance but pairs that are injected or created in the plasma annihilate away once they have cooled.

The model assumes a spherical geometry with the photons injected homogeneously throughout a spherical comptonizing cloud. As input source of soft photons \textsc{eqpair} includes a soft photon source modelled as a disc or single-temperature blackbody. Since the seed photon temperature is mainly decided by the maximum temperature disc photons, we use the blackbody option for simplicity. The part of the photons reaching us un-scattered is normally given in output by the model without the addition of any extra disc component. We have here however modified \textsc{eqpair} for use with a convolution reflection model and removed the blackbody from the output spectrum to avoid having the disc reflected in itself. We therefore model the direct disc emission with a separate disc component using \textsc{diskbb}. Following KD04, who investigated the temperature-luminosity correlation in spectra based on the same ASCA-PCA-HEXTE data as we use here, we assume that the disc is slightly truncated and using other more sophisticated disc models including relativistic effects thus makes no sense (but see \citealt{steiner} for a different view). The maximum temperature of the disc is allowed to differ from that of the input photons in \textsc{eqpair} since in reality we may see a somewhat cooler part of the disc in direct emission than that providing the seed photons.

\subsection{\textsc{RFXCONV} and the iron-line region}
Here, as in AHD13, we use the relatively new model {\sc rfxconv} to model reflection of the hard X-rays off the accretion disc. The {\sc rfxconv} model is a convolution model that combines the table models for the reflected spectra of \citet{2005MNRAS.358..211R}
from a constant density ionized disc, with the \textsc{ireflect} convolution variant of the \textsc{pexriv} Compton
reflection code by \citet{pexrav}. It includes a self-consistently calculated iron line, convolved with any continuum shape. The main parameters are the relative amplitude of the reflected component, $R$,
the inclination and the ionization parameter of the reprocessing
matter, $\xi$. For the inclination we use $i=70\deg$ \citep{orosz02}. We allow for relativistic smearing of the reflection features using {\sc kdblur}. 
We include photoelectric absorption in the form of \textsc{tbabs}. In AHD13 the fits to the PCA data were found to be
significantly improved by the addition of a narrow
absorption line (\textsc{gabs}) at $\sim6.8$ keV. Observations of similar strong
absorption lines have been reported in several high inclination
Galactic binary systems and are believed to represent iron K resonance
at 6.7 and 7.0 keV from outflowing material \citep[e.g.,][]{1998ApJ...492..782U, 2002ApJ...567.1102L, 2004ApJ...609..325U}. We thus include a line as \textsc{gabs} with central energy between 6.4 and 7.1 keV. 

Our initial single-component hybrid comptonization model of the form  \textsc{constant*tbabs*gabs(diskbb+kdblur*rfxconv*eqpair}. Compared to the similar HYB model in GD03 we use a broader energy range including the ASCA data, a better reflection model \textsc{rfxconv} and different assumptions about the compactness, see below. 
\section{Results}
\subsection{Single-component hybrid comptonization model for the 0.7--1000 keV continuum} 
Figure~1a shows the best fit total model of the form \textsc{constant*tbabs*gabs(discbb+kdblur*rfxconv*eqpair)}to the 0.7--1000 keV continuum data of XTE J1550-564. The parameters are listed in Table~1, left column (model \textsc{one-component hybrid}) and the components of the model are plotted in Fig.~1b. 

The spectrum contains a strong disc component. The best-fit inner disc temperature of the unscattered disc component is 0.56 keV, in agreement with or slightly higher than in DG03, KD04 and AHD13. We use the disc temperature as input for the seed photon temperature but initially allow them to be different. The best value for the seed temperature however stays the same as that of the inner disc and we thus set them equal in order to reduce the number of free parameters. 

\begin{figure*}
\begin{center}
\includegraphics[scale=0.3]{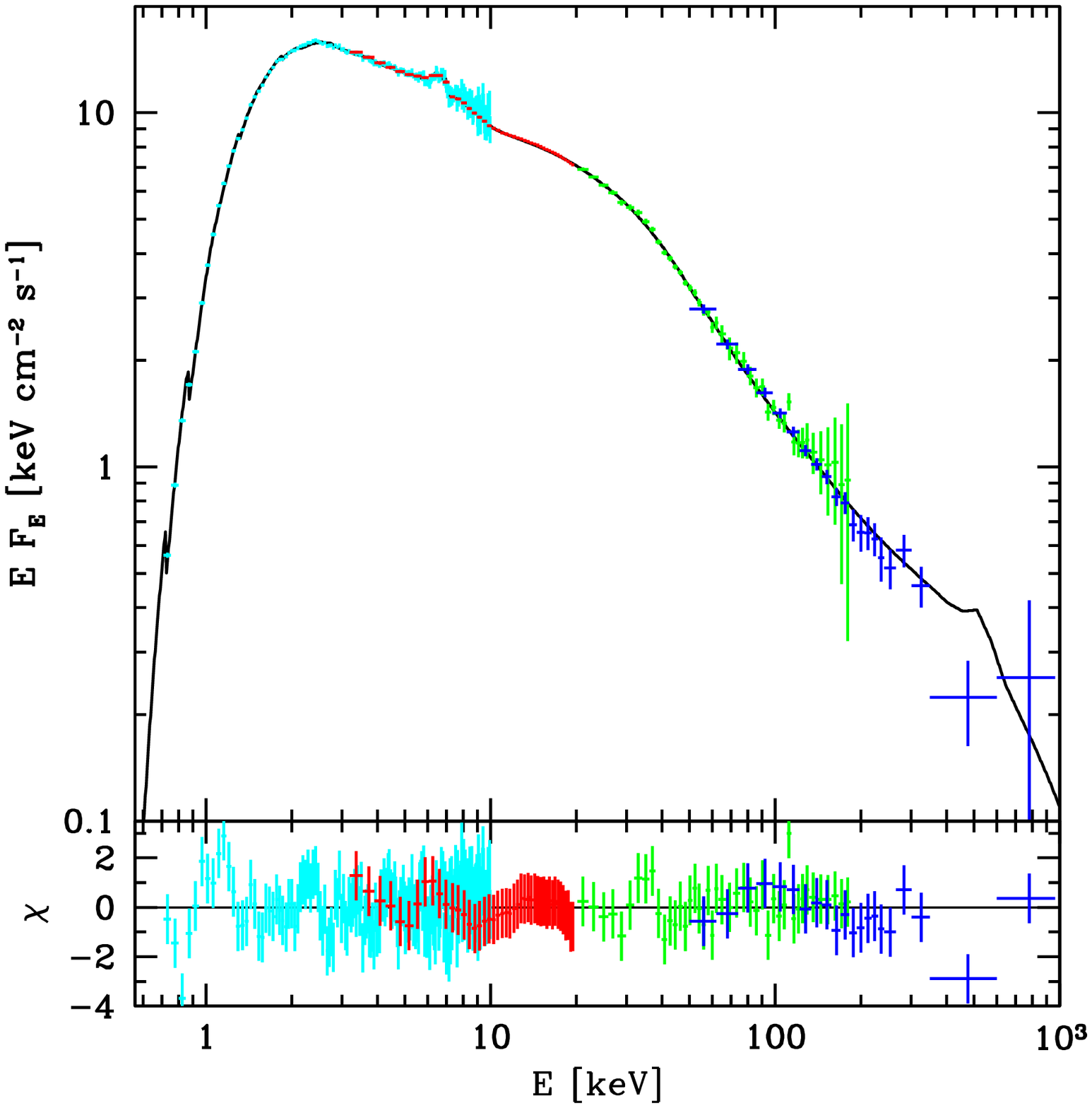}
\includegraphics[scale=0.3]{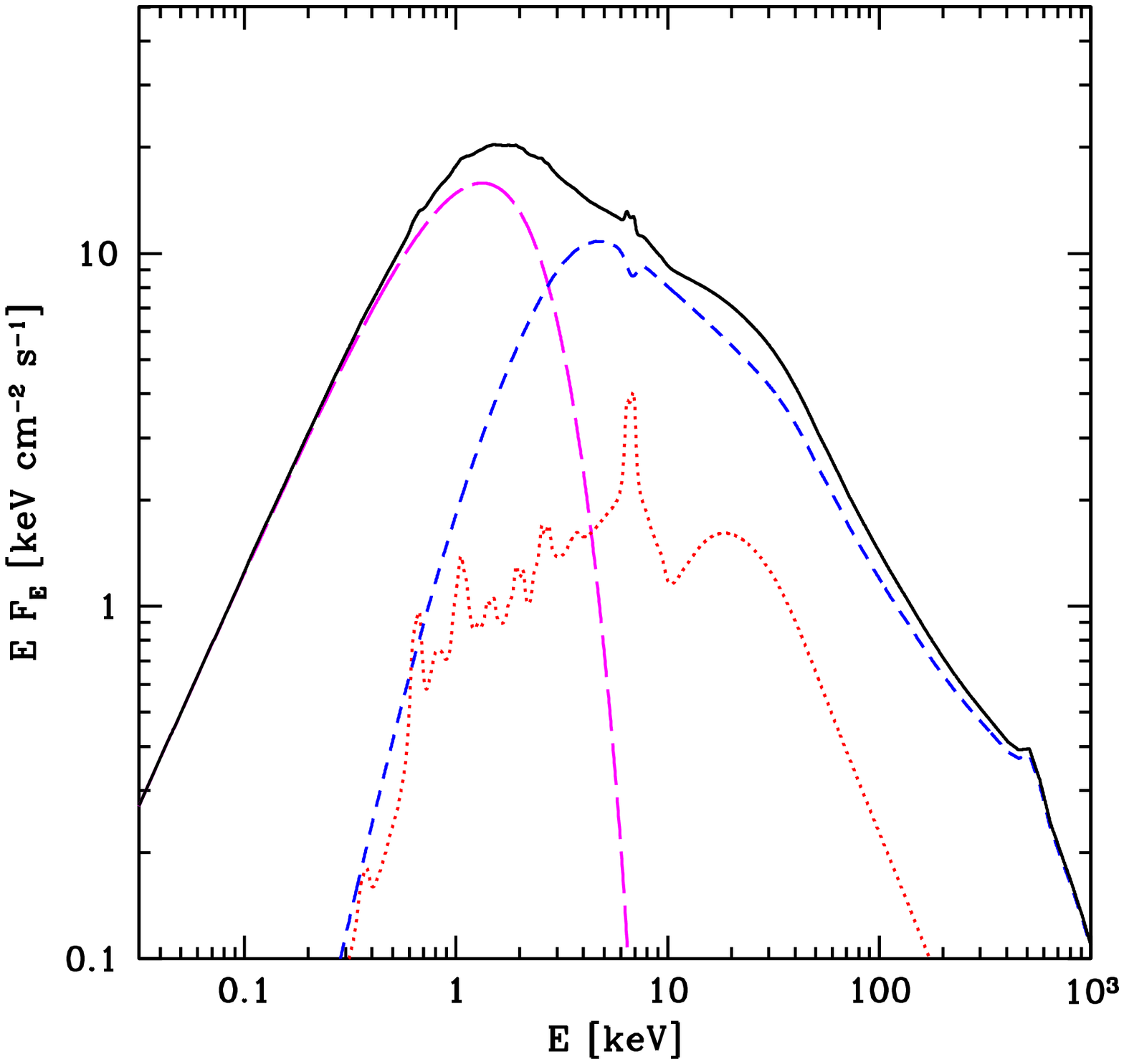}
\caption{One-component hybrid Comptonization model to the 0.7--1000 keV data of XTEJ1550-564. \textit{Left panel:} Data and model including residuals. ASCA data 0.7--10 keV in black, PCA 3--20 keV in red, HEXTE 20--200 keV in green and OSSE 50--1000 keV in blue. \textit{Right panel:} Components of the model. The unscattered blackbody in magenta long dashes, Comptonization from a hybrid electron distribution in blue short dashes and Compton reflection as red dots.} 
\end{center}
\label{onecomp}
\end{figure*}

We also see a very strong Comptonized component, typical for the very high state. Our best fit value of  \lh/\ls$\sim1$ indicates equal power in heating of the electrons as in cooling them off soft photons. The electron distribution is hybrid with $\sim50\%$ of the power to the electrons supplied as acceleration as opposed to thermal heating. The temperature of the low-end Maxwellian electrons in our best fit is $kT_{\rm e}=7.0$ keV. The optical depth is $\tau_{\rm T}=3.9$, and includes a small contribution ($\tau=0.1$) from pairs. We stress that in a soft spectrum with hybrid electrons present the rollover from the low energy Maxwellian distribution falls in the same energy range, $3kT_{e}\sim$20--30 keV, as the peak of the reflection bump and is therefore difficult to determine exactly. The exact values of the electron temperature and optical depth may therefore be somewhat degenerate within given errors. 

In our best fit, reflection is moderately high at 0.65 and comparable to values found in AHD13 and GD03. Reflection is highly ionized with best fit value $\log \xi$=2.7. The inclusion of a broad absorption line with $\sigma=0.40$ centered at 6.8 keV significantly improves the fit. We are not sure as to whether this line is real or if it is an artefact compensating for the strength of the fluorescent line as compared to continuum reflection in the \textsc{rfxconv} model. We stress that in \textsc{rfxconv} where the iron line is calculated self-consistently, it can not be arbitrarily moved in energy or strength (as when using e.g. \textsc{ireflect} + \textsc{gauss}). 

We include \textsc{kdblur} but find that the fit is insensitive to the value of the inner radius together with ionized reflection. We find  
that values lower than $R_{\rm in}=170$ are ruled out by the data and thus freeze the inner disc radius to this value rather than to a best-fit value even if $\chi^2$ tends to decrease for even larger values. We do not believe this value to represent a true lower limit on the inner disc radius. But it shows that a small inner radius with signs of relativistic effects on the reflection features is not compatible with the data if we assume ionized reflection. We find that the only way our data is compatible with small inner disc radius is if we assume zero ionization of the reflector. This does not only give a worse fit statistically, we also have good reason to believe that the reflection features should show signs of ionization since the disc is hot as well as strongly irradiated. The ionization parameter is rather a representation of an average than the true ionization state of the reflector as the reflection spectrum is a composite of many different parts of the vertical structure of the disc. Some contribution comes from the almost completely ionized surface while some comes from the almost completely neutral material further inside the disc, but very little reflection is expected from intermediate ionization states \citep{2000ApJ...537..833N, 2007MNRAS.377L..59D}.

\begin{table*}
\begin{tabular}{l l l l l}

MODELS & & {\tiny ONE-COMPONENT} & {\tiny TWO-COMPONENT} & {\tiny TWO-COMPONENT} \\
 & & {\tiny HYBRID} & {\tiny THERMAL + HYBRID} & {\tiny THERMAL + FULL NON-TH INJECTION} \\
\hline
$N_{\rm H}$ & $10^{22}$cm$^{-2}$ & 0.65$^{+0.02}_{-0.02}$ & 0.65$^{+0.02}_{-0.02}$ & 0.61$^{+0.01}_{-0.02}$ \\
E$_{\rm abs.line}$ & keV & $6.77^{+0.07}_{-0.06}$ & $6.77^{+0.08}_{-0.05}$& - \\
$\sigma_{\rm abs.line}$ & keV & $0.40^{+0.08}_{-0.07}$ & $0.40^{+0.12}_{-0.12}$ & - \\
Line depth$_{abs.line}$ & keV & $0.13^{+0.07}_{-0.06}$  & Ê$0.14^{+0.07}_{-0.08}$ & Ê$-$\\
$T_{\rm disc}$ & keV & 0.56$^{+0.14}_{-0.03}$ & $0.56^{+0.14}_{-0.04}$ & $0.56^{+0.01}_{-0.02}$ \\
\hline
$T_{\rm seed,var}$ & keV & - & $0.56f$ & $0.56f$ \\
$l_{\rm h}/l_{\rm s, var}$ & & - & $1.11^{+0.02}_{-0.04}$ & $1.11^{+0.02}_{-0.04}$ \\
k$T_{\rm e,var}^a$ & keV & - & $5.46$ & $5.46$ \\
$\tau_{\rm p,var}^a$ & & - & $6.48^{+0.73}_{-0.69}$ & $6.48^{+0.73}_{-0.69}$ \\
$\tau_{\rm T,var}^a$ & & - & 6.48 & 6.48 \\
Refl$_{\rm Êvar}$ & & - & $0.24^{+0.12}_{-0.11}$ & $0.24^{+0.12}_{-0.11}$ \\
\hline
T$_{\rm seed}$ & keV & $0.56f$ & $0.56f$ & $0.56f$ \\
$l_{\rm h}/l_{\rm s}$ & & 0.98$^{+0.04}_{-0.02}$ & $0.97^{+0.04}_{-0.04}$ & $0.89^{+0.08}_{-0.06}$ \\
$l_{\rm s}$ & & 500f & 500f & 500f \\
$l_{\rm nth}/l_{\rm h}$ & & 0.48$^{+0.07}_{-0.03}$ & $0.54^{+0.17}_{-0.05}$ & 1.0f \\
k$T_{\rm e}^a$ & keV & 7.00 & 6.96 & Ê5.49 \\
$\tau_p$ & & 3.77$^{+0.33}_{-0.17}$ & $3.59^{+0.24}_{-0.47}$ & $2.36^{+0.59}_{-0.24}$ \\
$\tau_T^a$ & & 3.93 & 3.77 & 2.51 \\
$\Gamma_{\rm inj}$ & & 2.27$^{+0.23}_{-0.17}$ & $2.45^{+0.55}_{-0.23}$ & $3.75^{+0.23}_{-0.43}$ \\
Refl & & 0.65$^{+0.13}_{-0.22}$ & $0.66^{+0.16}_{-0.24}$ & $1.0_{-0.21}$ \\
log $\xi$ & & 2.70$^{+0.09}_{-0.14}$ & $2.73^{+0.17}_{-0.09}$ & $3.47^{+0.06}_{-0.04}$ \\
$R_{\rm in}$ &r$_{\rm g}$ & $170{\rm f}^{b}$ & 170f & 170f\\

$\chi^2$/dof & & 245/290 & 240/291 & 327/295\\
\hline
\end{tabular}
\\
\begin{flushleft}
$^a$Calculated self-consistently by the model.\\
$^b$Lower limit\\
$^{\rm f}$Parameter fixed.
\end{flushleft}
\caption{Fit results of three different models for the continuum as described in Sections 5.1 and 5.3 . Parameters with subscript 'var' represent best-fit values for the thermal model for the spectrum of the fast variability as described in Section 4.2. Errors for these parameters are from the fit to the variability spectrum only. They were frozen to their best-fit values when determining the other parameters in the two-component models. The unabsorbed bolometric luminosity is $L_{tot}=3.7\times10^{38}$ for all models.}
\label{modeltable}
\end{table*}

\subsubsection{Compactness and annihilation lines} 
There is no sign of an annihilation line at 511 keV in the data of XTE J1550-564. The lack of observed annihilation features in GBH data is often used as motivation for constraining $l\rm_{s}\leq10$. While this may be justifiable in less luminous sources, like e.g. Cyg X-1 \citep{cygx1}, it is certainly not so in the more luminous sources and/or states. Our model gives a soft luminosity of $\sim1.7\times10^{38}$ erg s$^{-1}$. In the very high state, we expect a fairly large overlapping region of the hot flow and the disc. But a soft compactness of $\sim10$ would imply a comptonizing region extending beyond 500 $R_{\rm g}$ which is not easily reconciled with our current picture of the accretion geometry. More reasonable radii of the extension of the hot flow, 10--30 $R_{\rm g}$, give a range of \ls$\sim$200--700. Since the modelling is insensitive to this parameter (in all but the size of the annihilation line) we thus keep {\ls} frozen to the more realistic value of 500. Such a high compactness should lead to copious pair production given that enough high energy gamma rays are present. Using a realistic value for the compactness based on the luminosity together with good-quality high-energy data, we are thus in a position to constrain the injection spectrum of high energy electrons, determined by a combination of the power law index of injected electrons {\Gammainj} and their maximum gamma factor {\gammamax}. The parameter value for the maximum energy of the injected electrons {\gammamax} is insensitive to the fitting process and should be kept frozen. To determine allowed combinations of {\Gammainj} and  {\gammamax} we fix {\gammamax} to a range of different values between 3 and 1000 and calculate error ranges for {\Gammainj} for each {\gammamax}. Fig.~2 shows examples of different combinations of {\Gammainj} and {\gammamax} all giving acceptable fits ($\delchi^{2}/\nu \leq1$) to the broadband data, but with a preference for values around {\gammamax} $~\sim10$ and {\Gammainj} between 2.10--2.50 with best-fit value 2.27.

In addition, the OSSE data give an upper limit for the equivalent width EW of the line of 70 keV (at 3 $\sigma$ confidence). Fig.~3 shows best-fit models for different combinations of {\Gammainj} and {\gammamax} from Fig.~2. We find that all models with {\gammamax}$>$10 produce an annihilation line with EW$\geq$70 keV and can thus be ruled out based on the non-detection on basis of the OSSE data alone. We thus keep {\gammamax} frozen to 10 (since this parameter is rather insensitive to the fitting process as discussed above).

\begin{figure}
\begin{center}
\includegraphics[scale=0.7]{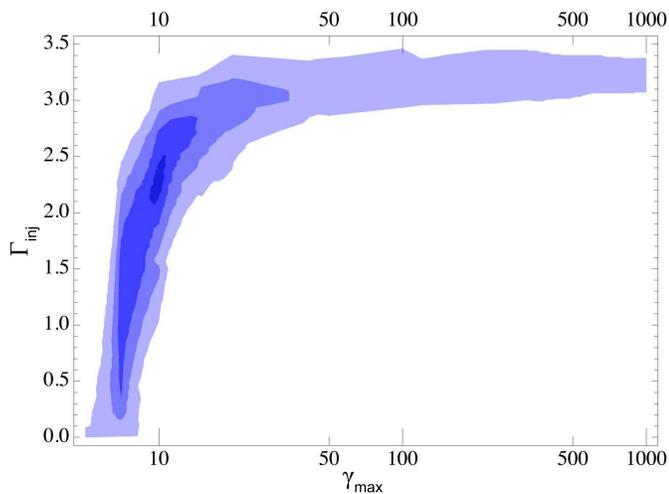}
\caption{Contour plot for $\Gamma_{\rm inj}$ versus $\gamma_{\rm max}$ from the one-component model fit to the data of XTE. Shaded regions show contours
corresponding to (from darker to lighter colour) 1$\sigma$, 3$\sigma$, 5$\sigma$ and 7$\sigma$.} 
\end{center}
\label{gammakorrar}
\end{figure}

\begin{figure}
\begin{center}
\includegraphics[scale=0.4]{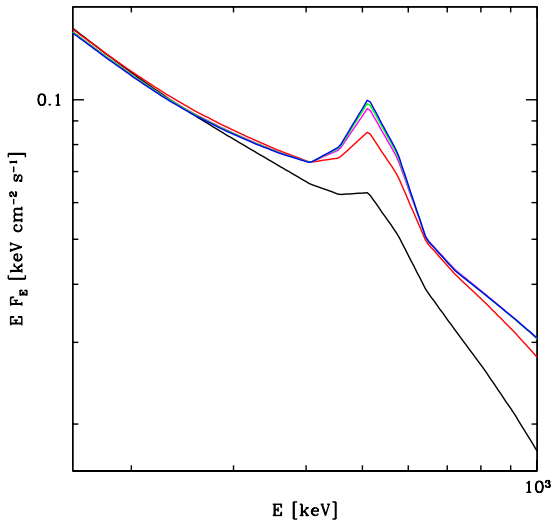}
\caption{Annihilation lines. Example best-fit models for different combinations of {\Gammainj} and {\gammamax}. From bottom to top \gammamax=10, 20, 50, 100 and 1000. Note that all these models give fits to the broadband data with $\chi^2$/dof $\leq 1$.} 
\end{center}
\label{lines}
\end{figure}

\subsection{The spectrum of the fast variability}
We now fit the spectrum of the 10--50 Hz variability for the co-added PCA observations. Since the data start only at 3 keV, we keep the disc temperature frozen to the best fit value from the continuum fit based on the ASCA data. There is no sign of a blackbody or disc component in the spectrum of the fast variability. This agrees with the findings in AHD13 and confirms the well known result that the fast variability does not come from the accretion disc in soft states (e.g. \citealt{churazov}). We thus keep the disc normalization frozen to 0. 

Even after removing the disc component the continuum model still does not give a good fit to the spectrum of the variability. It is clear that the two spectra have a different shape in the overlapping energy region (3--20 keV). The spectrum of the variability is harder with a higher \lh/\ls, meaning that the variable part of the flow sees less soft photons than the overall accretion flow. The optical depth is also higher. This result agrees with AHD13 who showed that the spectrum of the fast variability had a different shape, always harder than the continuum, in all PCA observations in the rise of the strong flare in 1998.

\begin{figure*}
\begin{center}
\includegraphics[scale=0.3]{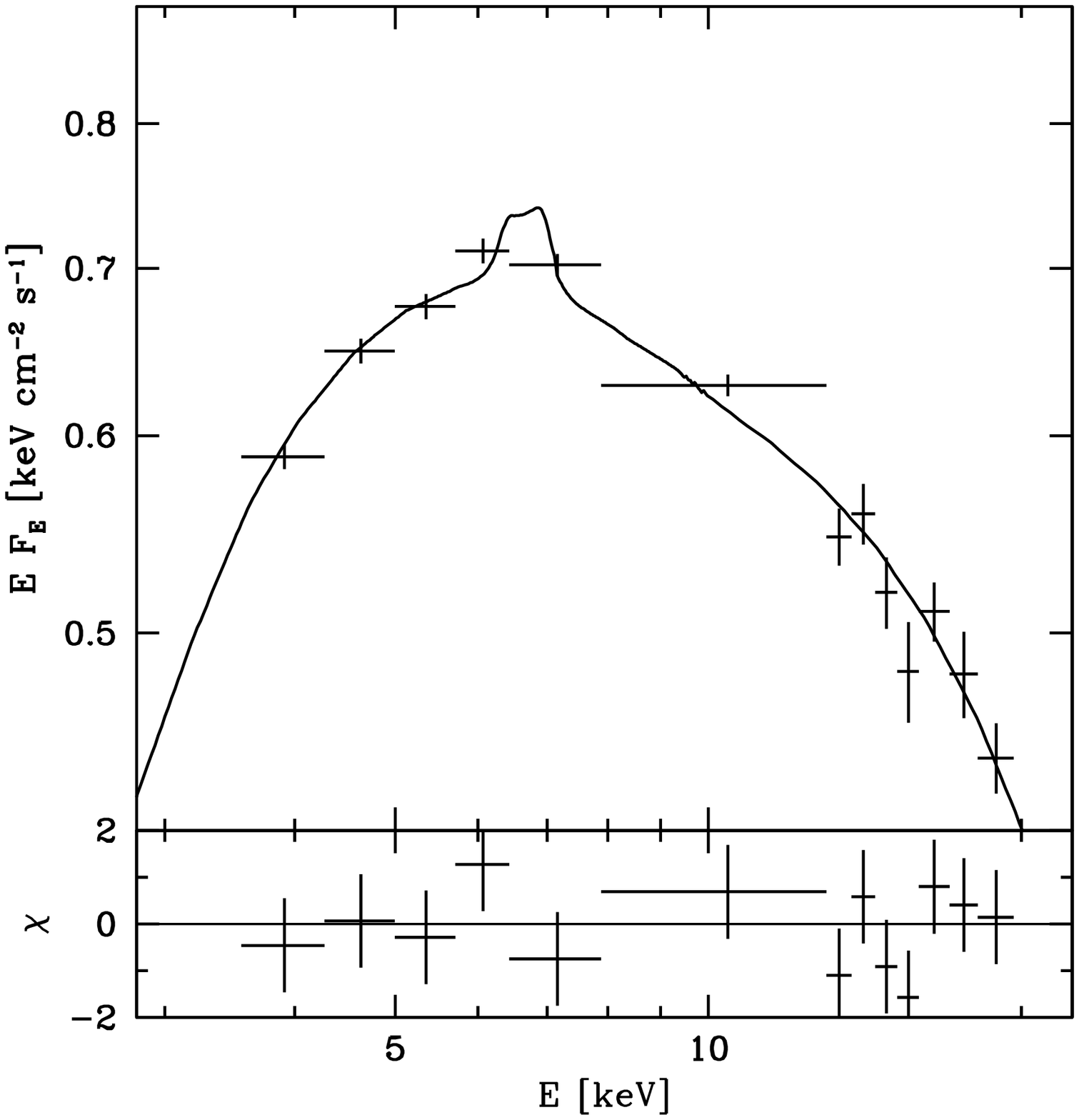}
\includegraphics[scale=0.3]{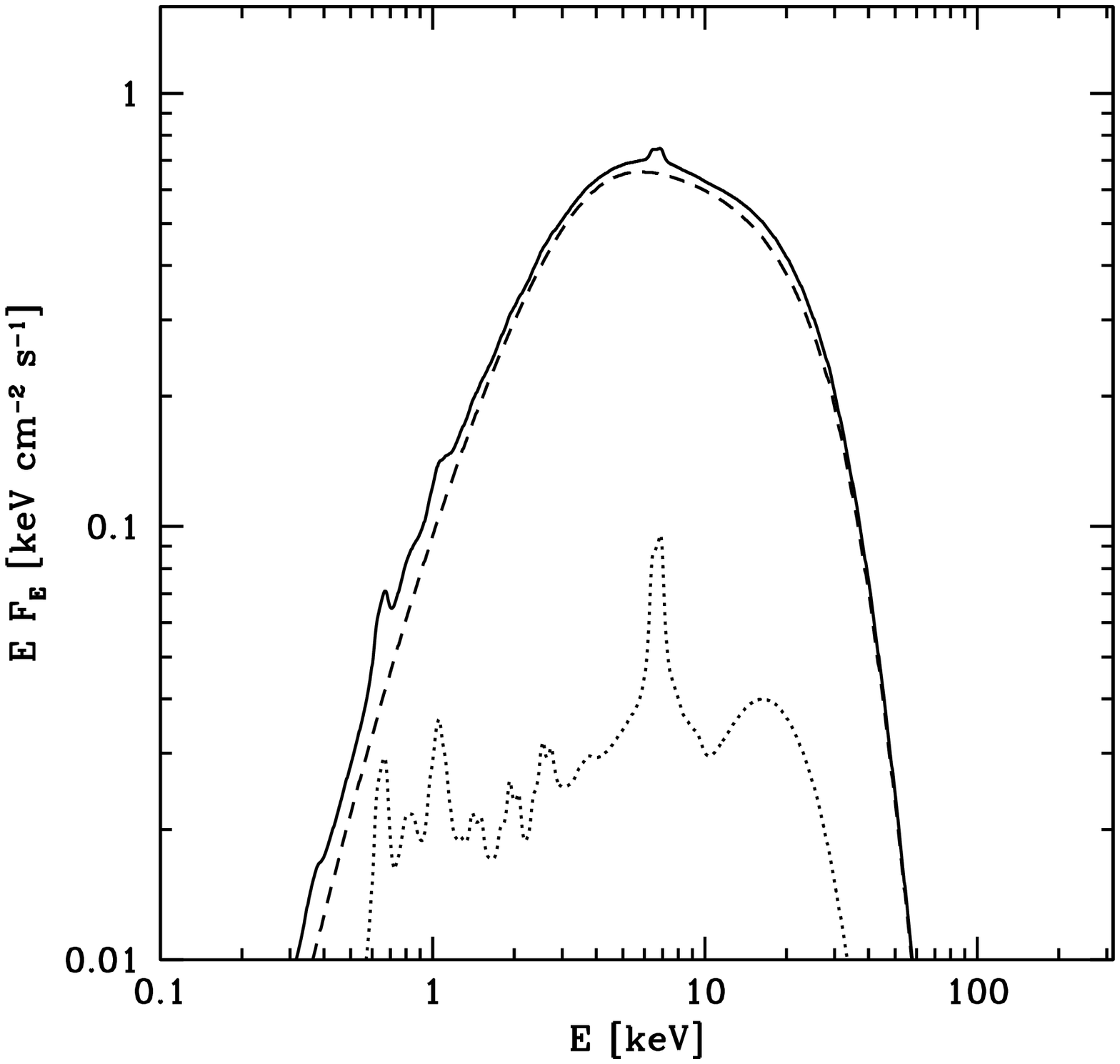}
\caption{Spectrum of the 50--100 Hz variability in XTEJ1550-564. \textit{Left panel:} The fourier resolved PCA data and best-fit (fully thermal) model including residuals. \textit{Right panel:} The model components are shown as dashed lines for the comptonized spectrum and dotted lines for its reflection.} 
\end{center}
\label{fourier}
\end{figure*}

Since our variability spectrum continues up to 20 keV only, we can thus not constrain the contribution of non-thermal comptonization or get a very good idea of the electron temperature from these data. We then have two obvious choices regarding the non-thermal electron distribution in the spectrum of the fast variability. One would be to freeze the ratio of non-thermal to thermal contribution and the parameters of the injection spectrum to those of the continuum, thus assuming the same non-thermal electron distribution in the variability spectrum. The other would be to freeze \lnth/\lh = 0 giving a purely thermal spectrum (as in AHD13, based on PCA data only). Since this choice does not strongly affect the spectral shape below 20 keV both variants should give a satisfactory ($\chisq/\nu \sim$1) fit to the data. We find that a pure thermal model gives a slightly better ($\chisq/\nu$=0.96 compared to 1.05) fit and stick to this model for simplicity. We stress that whether the electrons participating in the fast variability are in fact thermal or hybrid can only be determined by higher resolution data (in time and energy) at higher energies. Regardless of this choice, however, the spectrum of the variable component is considerably steeper than that of the time-averaged, meaning that even if there is non-thermal emission in the spectrum of the variability, it is not large compared to that of the time-averaged spectrum. 

In AHD13 it was found that all the variability spectra of the rising phase of the flaring of XTE J1550-564 had little or zero reflection, also this indicating that the variable photons are to a smaller degree intercepted by the disc. In our more detailed fit to the co-added frequency resolved spectrum we find that reflection, even if low in our best-fit (thermal) model, is not compatible with zero. Freezing $R=0$ only gives an acceptable fit if the absorption line is turned into an emission line at 6.45 keV, indicating the need for at least a fluorescent iron line even in the variability data (regardless of what is assumed regarding the electron distribution). Our best-fit thermal model to the spectrum of the fast variability has $R=0.24$, and requires no absorption line. For such low reflection, the ionization parameter can not be constrained. We therefore assume the same degree of ionization in the reflection of the variability spectrum as for the continuum. The best fit model and its components is shown in Fig.~4. The parameters are listed in Table~1 with subscript 'var' as part of the two-component models, see next section. (Please note that these values and their errors are from the fit to the variability spectrum only. In the two-component model fits they are kept frozen.)

\begin{figure*}
\begin{center}
\includegraphics[scale=0.3]{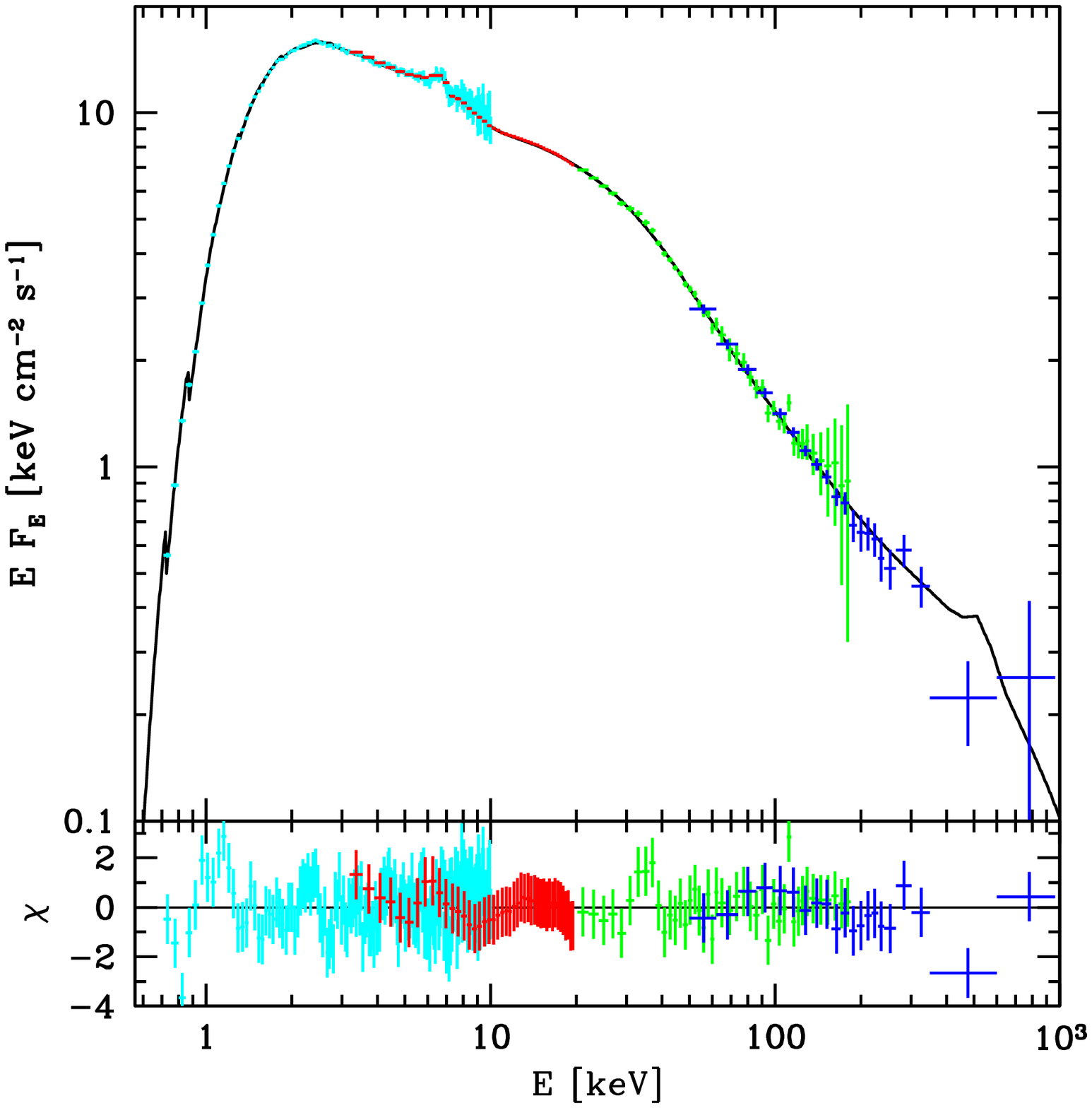}
\includegraphics[scale=0.3]{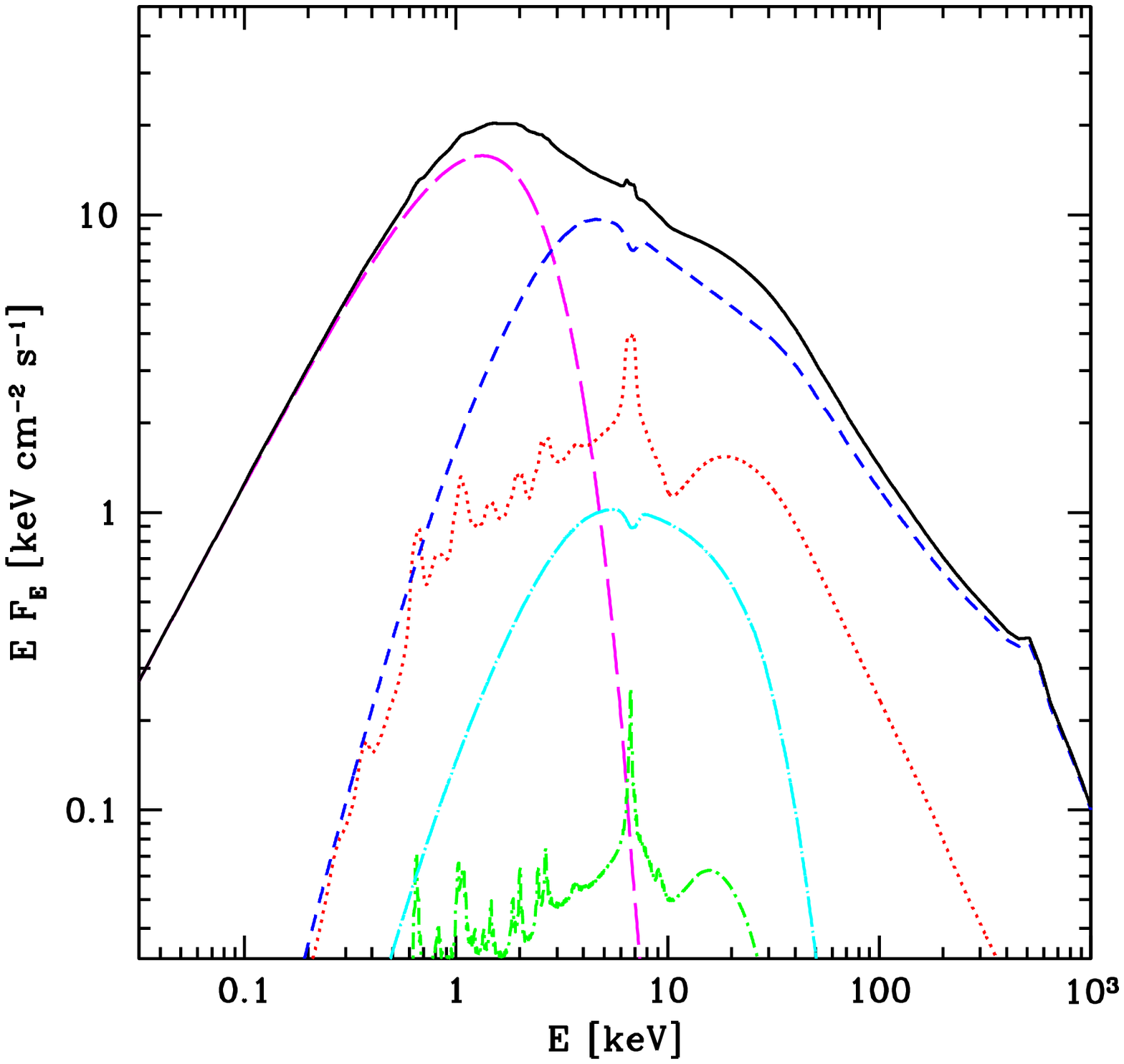}
\caption{Two-component thermal + hybrid Comptonization model to the 0.7--1000 keV data of XTEJ1550-564. One component is frozen in shape (but not in strength) to the best fit (fully thermal) model of the spectrum of the 50--100 Hz variability. \textit{Left panel:} Data and model including residuals. ASCA data 0.7--10 keV in black, PCA 3--20 keV in red, HEXTE 20--200 keV in green and OSSE 50--1000 keV in blue. \textit{Right panel:} Components of the model. The unscattered blackbody in magenta long dashes, the thermal Comptonization component matching the variability spectrum as cyan dot-dashes and its reflection in (shorter) green dot-dashes, Comptonization from the hybrid electron distribution in blue short dashes and its Compton reflection as red dots.} 
\end{center}
\label{twocomphyb}
\end{figure*}

\begin{figure*}
\begin{center}
\includegraphics[scale=0.3]{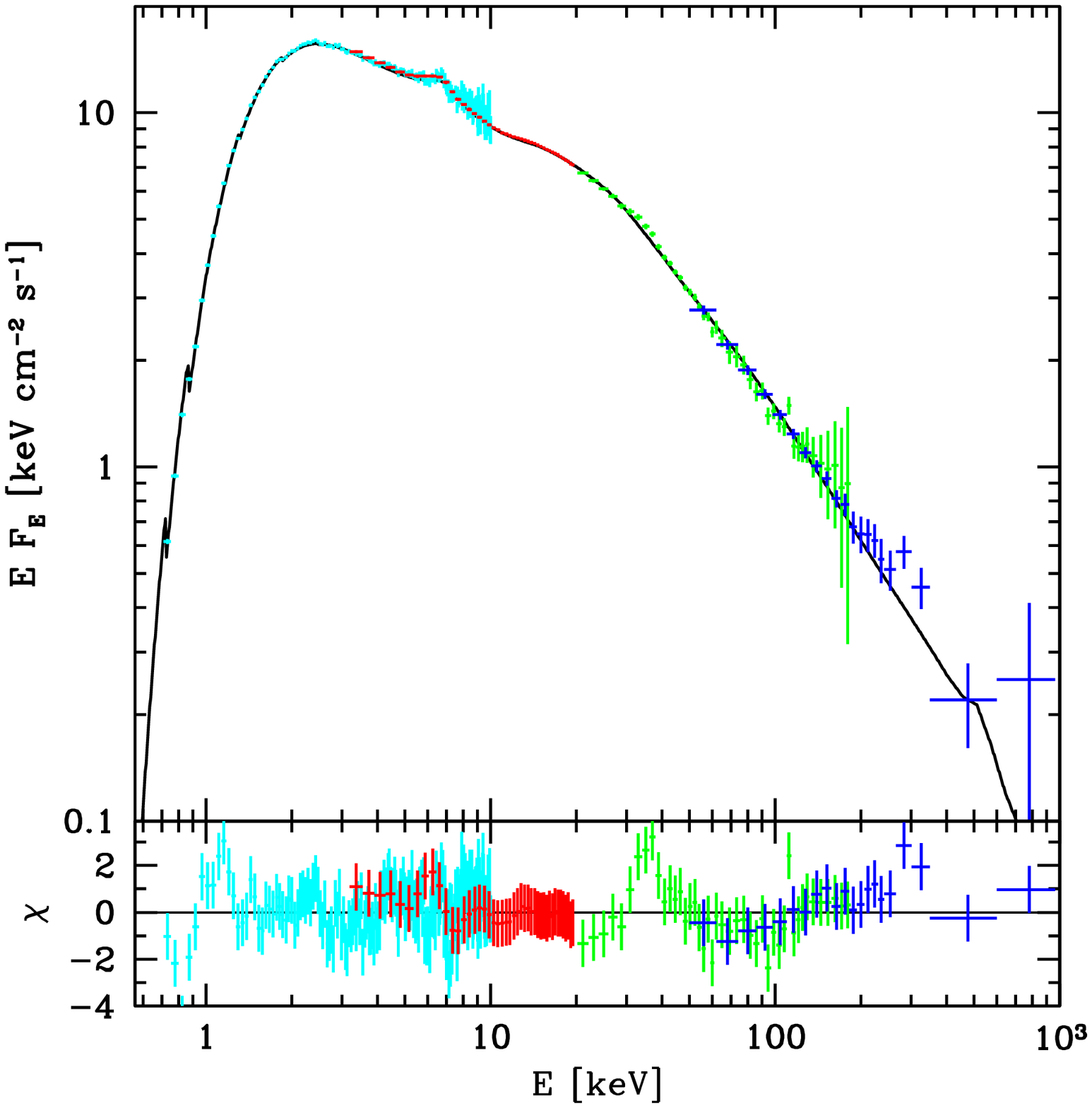}
\includegraphics[scale=0.3]{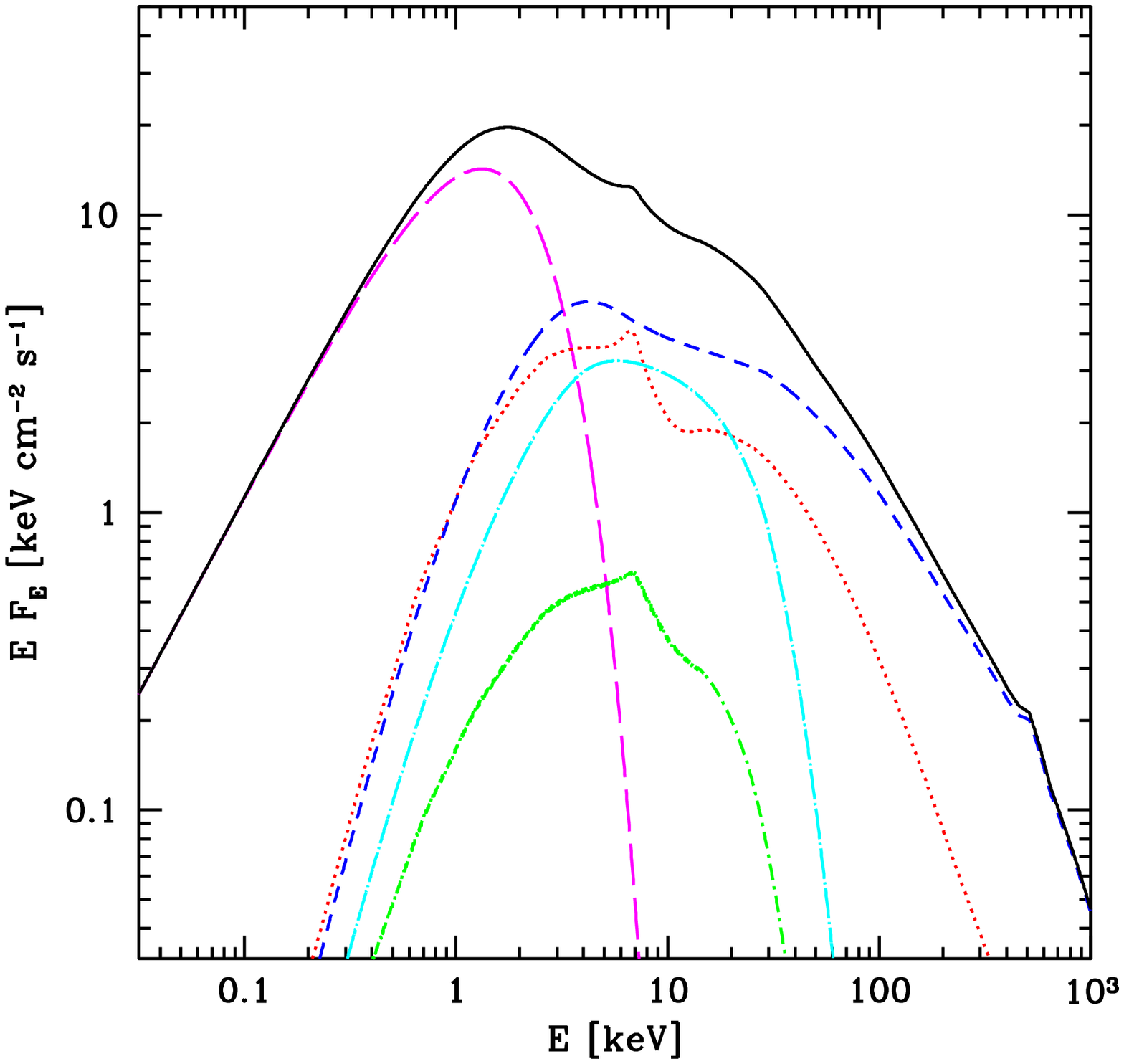}
\caption{Two-component thermal + hybrid Comptonization model with full non-thermal injection to the 0.7--1000 keV data of XTEJ1550-564. One component is frozen in shape (but not in strength) to the best fit (fully thermal) model of the spectrum of the 50--100 Hz variability. \textit{Left panel:} Data and model including residuals. ASCA data 0.7--10 keV in black, PCA 3--20 keV in red, HEXTE 20--200 keV in green and OSSE 50--1000 keV in blue. \textit{Right panel:} Components of the model. The unscattered blackbody in magenta long dashes, the thermal Comptonization component matching the variability spectrum as cyan dot-dashes and its reflection in (shorter) green dot-dashes, Comptonization from a hybrid electron distribution with purely non-thermal injection in blue short dashes and its Compton reflection as red dots. The ionization parameters of both reflection components are set equal and determined mostly by the stronger hybrid component.} 
\end{center}
\label{twocompnth}
\end{figure*}

\subsection{Two-component comptonization models}
Regardless of the details of the electron distribution and exact strength of reflection, the fact that the spectrum of the fast variability differs from that of the continuum already below 20 keV suggests that that there are at least two spatially separated electron distributions responsible for the comptonized emission of the total spectrum. This means that the flow is inhomogeneous and the total continuum should be the sum of the variable and non variable (on short timescales) part(s). A more physically correct broadband spectral model should thus contain at least two Comptonization components. This was already briefly discussed in AHD13 although the data used there (only PCA 3--20 keV) did not allow for any exact modelling. In DG03 a two-component model consisting of a hybrid plus an additional thermal component was actually found to give the best fit to the data (their model HYBTH). The extra component in that paper was however ambiguous and as argued there may have been only something artificial compensating for the not-so-correct reflection model. With the help of the fast variability we are here able to show that an additional component should in fact be present in the broadband spectrum. 

\subsubsection{Two component thermal+hybrid model}
We now fit the spectrum with a model consisting of a disc blackbody+ one thermal Comptonization component representing the fast variability in addition to the hybrid component representing the part of the comptonized spectrum that is not variable on fast timescales. We model both Comptonized components with eqpair and the total model is thus \textsc{constant*tbabs*gabs(rfxconv*eqpair + kdblur*rfxconv*eqpair)}. The parameters of the additional thermal component are frozen to that of the model for the variability spectrum, except for the normalization that is a free parameter and allowed larger than or equal to that of the variability spectrum which we use as input value. (Keeping it frozen at the same value would be to assume that this component is 100 per cent variable). 

To start with, we use the parameter values from the one-component model as input values for the parameters of the 'stable' hybrid component. It turns out that the inclusion of the additional component does not make large changes to the overall spectrum which is still dominated by the hybrid component, the parameters of which are very similar in the two models. The normalization of the additional thermal component is only slightly higher than the minimum required by the variability spectrum. Besides from it being weak, this also indicates that this additional component is indeed highly variable. The best-fit two-component model is shown in Fig.~5 and its parameters given in Table 1, column 2. Since the fit is very sensitive to the input values, we also try to use a higher normalization as input for the thermal component. We find that we get still acceptable ($\chisq/dof <1.0$) but slightly worse fits for the thermal component being a factor of $\sim$3 higher (corresponding to it being stronger but less variable). The addition of the extra component does not alter the limits on \Gammainj\ and \gammamax\ from the one-component model.

\subsubsection{Two component model with pure nonthermal injection}
To try to separate the thermal and non-thermal contributions, we also model the spectrum with \lnth/\lh\ fixed to 1.0 for the 'stable' hybrid component. This represents full non-thermal injection i.e. that the power supplied in terms of injection of high energy electrons is the sole energy supply to this part of the spectrum. This would be expected if this emission is powered by e.g. magnetic flares above the disc. (Note that this will not result in a pure non-thermal spectrum since the low end will always thermalize. Modelling non-thermal injection dominated spectra as pure power laws is thus not fully physically motivated even if such a model may give a good statistical fit, see e.g. model PLTH in GD03 and their discussion). The result is shown in Fig.~6 and best-fit parameters for this model are listed in Table 1, column 3. In this model the thermal component, matching the variability spectrum, and the remaining 'stable' non-thermal component are of comparable strength. The model requires very high strongly ionized reflection and a steep non-thermal injection spectrum, regardless of \gammamax, but as before, \gammamax$\ge10$ overestimates the annihilation line and gives a line flux above the detection limit for OSSE. This model does however not give a very good fit to the data ($\chisq/dof$ =1.11, compared to 0.84 for the one-component model and 0.82 for the two-component thermal+hybrid model). It would thus seem that the part of the spectrum not variable on short timescales is not compatible with being powered solely by non-thermal processes but that there are also important contributions from direct heating.

\section{Discussion}

\begin{figure}
\begin{center}
\includegraphics[scale=0.5]{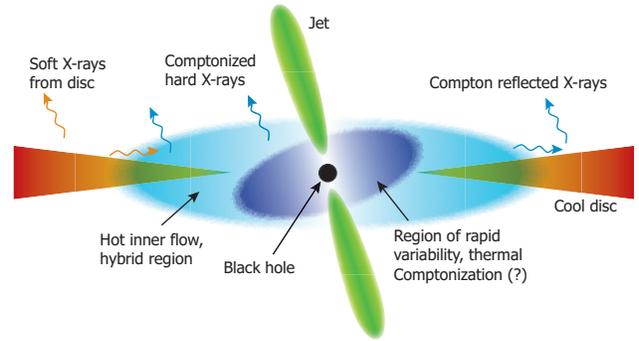}
\caption{Envisaged geometry of the accretion flow with at least two Comptonizing regions, one with rapid variability, possibly with a purely thermal electron distribution, close to the black hole. The rest of the flow which shows strong signs of containing both thermal and non-thermal electrons, may in turn be made up of several spatially separated regions with different properties and electron distributions.} 
\end{center}
\label{geometry}
\end{figure}

\subsection{Limits on pair-production}
Despite being predicted by most models, no annihilation lines are actually observed in the spectra of GBHs (except for the claimed detection in 1E 1740.7-2942, \citealt{1991ApJ...383L..49S}). This can only have three explanations: either no pairs are produced, they do not annihilate or they do annihilate but are hot when they do causing doppler-smearing of the line making it non-detectable. For luminous GBHs the compactness should be high, rather several hundred (see Section 4) than $\sim10$ that is commonly used to avoid the problem of creating strong annihilation lines not seen in the data. Pair-creation rates are thus very high. With so many soft photons present, Compton cooling is rapid and the pairs should lose most of their energy before they annihilate, so we do not expect broadening of the line. Since the density of pairs is high, created pairs cannot be accreted or escape from the source before they annihilate. The only reasonable explanation is thus a deficit of high energy electrons capable of up-scattering photons above the pair production threshold. This would require either a very steep injection spectrum or a low maximum Lorenz factor of the injected electrons, or both. With good quality high energy data, like that used here, the the slope of the steady-state electron distribution can be rather well constrained and we can determine limits on combinations of {\Gammainj} and {\gammamax}. The non-detection of the annihilation line in this data limits {\gammamax} to $\leq10$ and thus {\Gammainj} further to 2.10--2.50 in our best-fit models (the one-component and the two-component hybrid model). Assuming full non-thermal injection (\lnth/\lh=1, two-component non-thermal model) requires steeper injection \Gammainj = 3.4--4.0, but this model gives a worse fit to the data.

Our results suggest that pair production may not be as important in GHBs as previously assumed, at least not in the very high state and that preferred acceleration mechanisms do not need to produce many electrons above \gammamax=10.

\subsection{Multi-zone comptonization and geometry of the flow}
It is by now widely accepted that the origin of the fast variability in the emission from GBHs in the soft and very high state is not the accretion disc but rather the hot flow or corona. Our results, here and in AHD13, that the frequency resolved spectrum of the 10--50 Hz variability does not contain any disc component confirms this picture. Further, the fast variability is generally assumed to arise in a region very close to the black hole. Our results, here and in AHD13, that the variability spectrum shows less reflection and has a harder slope than the continuum indeed suggest that the variable region intercepts less soft photons and is located far away from the accretion disc. The study of the spectral evolution of the variability spectrum versus the continuum in AHD13 further showed that the spectrum of the variability is less sensitive to changes in the accretion rate (and thus presumably changes in the inner disc radius) than the overall continuum spectrum. Thus, the spectrum of the fast variability and its evolution seem to agree with the origin of the fast variability being a hot flow close to the black hole far away from the disc.The envisaged geometry is shown in Fig.~\ref{geometry}. For a physical model for how to create the observed broad-band variability in the very high state, including the QPO arising in a hot flow, see \citet{ing11, ing12}.

Since the variability data covers energies up to 20 keV only, the temperature and electron distribution in the variable region can not be constrained but is consistent with being purely thermal. To confirm or disprove this we need to look at frequency resolved spectra at higher energies. Such data are at present not available but may be provided by the Hard X-ray Telescope (HXT) aboard the planned ASTRO-H mission. The steeper spectrum of the variable component, however, already shows that most of the observed non-thermal emission is produced not in the innermost variable region, but further out in the flow. It has been suggested that the non-thermal emission in accretion flow of GBHs are the result of high-energy electrons accelerated in magnetic reconnection flares above the disc surface. If this is the case then the non-thermal emission arises much closer to the disc and should be softer and show stronger signatures of reflection than the variable emission. In our two-component models, the `stable' hybrid component is both softer and has higher reflection than the variability component. We find that this stable component is not likely to be powered by purely non-thermal injection but requires half of the energy to be supplied as direct heating. This is however still compatible with the non-thermal emission being produced by magnetic flares above the (truncated) disc, since there can also be 'contaminating' thermal plasma from the outer regions of the flow.
In \textsc{eqpair} thermalization is due to Coloumb interactions only. An even more important thermalization process, not included in the code, is synchrotron self-absorption. The importance of this effect for the emission from compact sources was first pointed out by \citet{synchboiler} and has been further investigated by e.g. \citet{Poutanen09} and \citet{Malzac09}. In the presence of a magnetic field the electron distribution may indeed appear thermal even if the original acceleration mechanism would have produced a non-thermal distribution. It is also quite possible that the hybrid part of the flow is in fact a combination of several regions, each with a different electron distribution.

We have here investigated the very high state. The situation is likely to be different in the hard or the more classical soft state. In the classical soft state (of this and other sources) variability is usually strongly suppressed. This is consistent with the innermost hot flow collapsing into an accretion disc extending all the way to the innermost stable orbit. The extra component associated with the fast variability is thus not expected to be present in the classical soft state. In the hard state, variability is stronger and the component matching the variability may play a more important role or even dominate the spectrum. In AHD13 it was shown that for the harder of the very high state spectra of XTE J1550 the difference between the variability spectrum and the total continuum was indeed less than for the softer spectra. Unified spectral and timing studies of Cyg X-1 by \citet{yam13} have also shown the need for two Comptonized components in the hard state, one of which seem to be connected with the fast variability.

\section{Conclusions}
Galactic black holes are luminous objects and with data covering most of the soft X-rays from the accretion disc, the absolute soft compactness can be calculated, and it is high $>100$. The lack of evidence of any annihilation lines in their spectra should thus tell us that perhaps there are less high-energy electrons present in the flow than usually assumed. We have analysed the broadband 1--1000 keV spectrum of the GBH XTE J1550-564. Using a realistic value for the compactness we have calculated limits on the high energy electron distribution and find that the slope of the OSSE data constrains the mean electron energy to be $\gamma \sim$ a few, requiring either a steep electron spectrum and/or a low maximum electron energy. The lack of an observable annihilation line favours \gammamax $\leq$10 and \Gammainj=2.1--2.5. 

We also model the frequency resolved spectrum of the 10--50 Hz variability from this source and find that it differs from the continuum, not only in that it contains no sign of a disc and has very low reflection. It also has a different shape of the comptonized part of the spectrum. This means that the comptonized flow itself is inhomogeneous and its total spectrum must be made up by at least two components, one variable on short timescales and one not, where the continuum should be a sum of these. In our interpretation the fast variability originates in the innermost parts of the accretion flow close to the black hole and far away from the accretion disc. This region is spatially and physically separated from the origin of the rest of the emission which may in turn be a sum of several smaller regions. 

We find that the spectrum of the fast variability is consistent with being fully thermal. This could be confirmed or disproved with access to better quality high-energy data from e.g. the HXT onboard ASTRO-H, but our results already require that the observed non-thermal emission is produced predominantly further out in the flow. However, the time-averaged 
emission is not consistent with being fully non-thermal, indicating that direct heating of the electrons is also still important further out in the flow.

\section*{Acknowledgments}
This work was supported by the Wenner-Gren Foundations (LH) and The Royal Swedish Academy of Sciences (MA). We used data obtained through the High Energy Astrophysics Science Archive 
Research Center (HEASARC) Online Service, provided by NASA/Goddard Space Flight Center.

\end{document}